# Octet Magnetic Moments and their Sum-Rules in Statistical Model

*M. Batra[1], A. Upadhyay[1]
[1]SPMS, Thapar University, Patiala-147201,INDIA
* email: mbatra310@gmail.com, alka@thapar.edu

## Introduction

The analysis of deep-inelastic DIS scattering experiments and hyperonic β-decays of octet baryons have brought forward evidence for the relation of magnetic moments with spin-polarization densities of octet baryons. This makes us clear that octet baryonic magnetic moments may get sufficient contribution from sea-quarks and gluons. Although simple quark model provides good results for magnetic moment of proton and neutron, still more accurate measurements of nucleon may differ from SQM predictions by 0.2%. The present work analyzes the contribution of sea-quarks to magnetic moment of all octet baryons. For this, the methodology is based on the statistical model presented in ref. [1] and [2].

## Theoretical Framework

The statistical model had been successfully implemented to explain the magnetic moments and other spin-related properties of a nucleonic system [1-2]. The model is based on assumption of hadrons as quark-gluon Fock states so that each Fock state shares some part of total probability associated with quark-gluon Fock states. The Fock states include different sub-processes like $|q\bar{q}g\rangle, |q\bar{q}gg\rangle, |q\bar{q}q\bar{q}\rangle$ etc. The methodology is based on defining a suitable wave-function [1] in such a way that each term in the wave-function contains suitable combinations of valence and sea so as to maintain the anti-symmetric and spin ½ nature of baryonic system.

$$\left|\Phi_{\frac{1}{2}}^{\uparrow}\right\rangle = \frac{1}{N}[\phi_1^{(\frac{1}{2})^{\uparrow}} H_0 G_1 + a_8 \phi_8^{(\frac{1}{2})^{\uparrow}} H_0 G_8 + a_{10} \phi_{10}^{(\frac{1}{2})^{\uparrow}} H_0 G_{\overline{10}}$$
$$+ b_1[\phi_1^{\frac{1}{2}} \otimes H_1]^{\uparrow} G_1 + b_8 (\phi_8^{\frac{1}{2}} \otimes H_1)^{\uparrow} G_8 + b_{10} (\phi_{10}^{\frac{1}{2}} \otimes H_1)^{\uparrow} G_{\overline{10}}$$
$$+ c_8 (\phi_8^{\frac{3}{2}} \otimes H_1)^{\uparrow} G_8 + d_8 (\phi_8^{\frac{3}{2}} \otimes H_2)^{\uparrow} G_8]$$

Each term is associated with a co-efficient which signifies the relative probability of quark-gluon Fock states with spin and color quantum numbers. The sea contributions has been categorized as scalar with coefficients ($a_0, a_8, a_{\overline{10}}$), vector with coefficients ($b_1, b_8, b_{\overline{10}}, c_8$) and tensor ($d_8$). The scalar sea represents coupling of spin $\frac{1}{2} - q^3$ coupled to spin 0 sea while vector and tensor sea represents coupling of valence with spin 1 and spin 2 sea. The estimation of all the coefficients is our main concern because each co-efficient signifies the probability associated with different Fock states. To calculate the coefficients, the first step is to use principle of detailed balance [2,3] in order to find the flavor probability of each Fock state. The principle assumes the balancing of any of two Fock states with each other. This leads to the conclusion that ratio of transition rates between any two of Fock states is equal to ratio of their densities.

$$|A\rangle \xrightleftharpoons[R_{B\to A}]{R_{A\to B}} |B\rangle \text{ and } \frac{R_A}{R_B} = \frac{\rho_A}{\rho_B}.$$

Various subprocesses included in the transition processes are: $g \rightleftharpoons gg, q\bar{q} \rightleftharpoons g, q \rightleftharpoons qg$. The generalized expressions for the rates of transitions between any two processes can be found as:

1). When both the processes $g \rightleftharpoons gg$, $q \rightleftharpoons qg$ are included:
$$\frac{\rho_{i,j,l,k}}{\rho_{i,j,l,k-1}} = \frac{(3+2i+2j+2l+k-1)}{(3+2i+2j+2l)k + \frac{k(k-1)}{2}}$$

Where i refer to $\bar{u}u$ pairs, j refer to $\bar{d}d$, l refers to $\bar{s}s$ and k refers to no. of gluons.

2). When the processes $g \rightleftharpoons s\bar{s}$ are included:

The non-negligible mass of strange quark limits the exchanges between the gluons and strange quark anti-quark pair. Gluon must possess the free energy at least greater than the two times the mass of strange quark. Using free energy distributions for gluons, final expression for transition rates between gluon and strange quark anti-quark pair can be derived as:

$$\frac{\rho_{i,j,l,k}}{\rho_{i,j,l+k,0}} = \frac{(k(k-1)\ldots 1(1-C_0)^{n-2l-1}(1-C_1)^{n-2l}\ldots(1-C_{l-1})^{n+k-2})}{(l+1)(l+2)\ldots(l+k)(l+k+1)}$$

All expressions in terms of $\rho_{0,0,0,0}$ are mentioned below:

| Baryon | Expression |
|---|---|
| $\Sigma^+$ | $\dfrac{\rho_{i,j,l+k,0}}{\rho_{0,0,0,0}} = \dfrac{2}{i!\,i+1!\,j!(j+1)!(l+k)!(l+k+1)!}$ |

| | | | |
|---|---|---|---|
| $\Sigma^-$ | $\dfrac{\rho_{i,j,l+k,0}}{\rho_{0,0,0,0}} = \dfrac{2}{i!i!j!(j+2)!(l+k)!(l+k+1)!}$ | | |
| $\Sigma^0$, $\Lambda^0$ | $\dfrac{\rho_{i,j,l+k,0}}{\rho_{0,0,0,0}} = \dfrac{1}{i!(i+1)!j!(j+1)!(l+k)!(l+k+1)!}$ | | |
| $\Xi^0$ | $\dfrac{\rho_{i,j,l+k,0}}{\rho_{0,0,0,0}} = \dfrac{2}{i!i+1!j!(j)!(l+k)!(l+k+2)!}$ | | |
| $\Xi^-$ | $\dfrac{\rho_{i,j,l+k,0}}{\rho_{0,0,0,0}} = \dfrac{2}{i!i!j!(j+1)!(l+k)!(l+k+2)!}$ | | |

These probabilities differ for each of the baryonic system due to their quark content. The above expression gives the probabilities in the form of ratios. The normalization condition $\Sigma_{i,j,k,l} \rho_{i,j,k,l} = 1$ gives the individual probabilities.

The statistical model is based on computation of individual multiplicities denoted in the form of various ratios. This ratio is in the form of $\rho_{j_1,j_2}$ where the core quark part carry spin or color $j_1$ and the gluons carry $j_2$ (spin or color) so that resultant probability for $j_1$ and $j_2$ is ½. The details of the multiplicities are given in ref. [1, 2]. These multiplicities face a normalization procedure in order to calculate the values of co-efficients ($a_0, a_8, a_{\overline{10}}, b_1, b_8, b_{\overline{10}}, c_8, d_8$). These coefficients provide us the contribution of sea. In more suitable way, two parameters $\alpha$ and $\beta$ can be defined so that we can relate octet magnetic moments in terms of these parameters only.

$$\alpha = \frac{2(6 + 3a_8^2 - 2b_1^2 - b_8^2 + 4b_8 c_8 + 5c_8^2 - 3d_8^2)}{27(1 + a_{10}^2 + a_8^2 + b_1^2 + b_{10}^2 + b_8^2 + c_8^2 + d_8^2)}$$

$$\beta = \frac{3 - 9a_{10}^2 - 3a_8^2 - b_1^2 + 3b_{10}^2 + b_8^2 + 8b_8 c_8 - 5c_8^2 + 3d_8^2}{27(1 + a_{10}^2 + a_8^2 + b_1^2 + b_{10}^2 + b_8^2 + c_8^2 + d_8^2)}$$

**Magnetic Moments of lowest lying $J^P = 1/2^+$ Baryonic States:**

Computation of magnetic moments of all spins ½ baryons depend upon the above defined the two key parameters. For this, one should have a suitable expression for calculation of magnetic moment of each baryon. To find the relevant expressions, we apply eigen operator for the magnetic moments $\hat{\mu} = \sum_q \dfrac{e_q^i}{2m} \sigma_z^q, q = (u,d,s)$ giving rise to the values dependent of α, β and $\mu_q$'s. The table below shows the values of calculated magnetic moments of all baryons:

| Magnetic Moment | Expression | Statistical Model | Expt. Result or PDG[31] |
|---|---|---|---|
| $\mu_{\Sigma^+}$ | $3(\mu_u \alpha - \mu_s \beta)$ | 2.374 | 2.458±0.01 |
| $\mu_{\Sigma^0}$ | $\dfrac{3}{2}(\mu_u \alpha + \mu_d \alpha - 2\mu_s \beta)$ | 0.775 | ----- |
| $\mu_{\Sigma^-}$ | $3(\mu_d \alpha - \mu_s \beta)$ | -1.11 | –1.16±0.025 |
| $\mu_{\Xi^0}$ | $3(\mu_s \alpha - \mu_u \beta)$ | -1.48 | –1.25±0.014 |
| $\mu_{\Xi^-}$ | $3(\mu_s \alpha - \mu_d \beta)$ | -0.51 | –0.65±0.0025 |
| $\mu_{\Sigma\Lambda}$ | $-\dfrac{\sqrt{3}}{2}(\alpha + 2\beta)(\mu_u - \mu_d)$ | -1.62 | 1.61±0.08 |

**Results and Conclusion:**

We apply statistical method to find the contribution of all types of Fock states to magnetic moment of all octet baryons. We find that the total magnetic moment of all octet baryons is matching well with the experimental data. In addition to this, we can also check the percentage of individual contributions in the form of scalar, vector and tensor sea from the coefficients for each baryonic system. Moreover, the latest studies show that the total magnetic moment of baryons also get remarkable contribution from the angular excitations and symmetry breaking corrections in SU(3). Although the present work calculates the magnetic moments for octet baryons where strange mass corrections appear in sea only but this model can be modified accordingly to check SU(3) symmetry breaking for valence quarks too. The well-known sum rules for magnetic moments of all octet baryons are GMO and CG sum rules. These sum rules are also checked in our model.

$$S_{CG} = \mu_p - \mu_n - (\mu_{\Sigma^+} - \mu_{\Sigma^-} - \mu_{\Xi^0} + \mu_{\Xi^-}) = -0.67$$

$$S_{GO} = 6\mu_\Lambda + \mu_{\Sigma^+} + \mu_{\Sigma^-} - 2(\mu_p + \mu_n + \mu_{\Xi^0} + \mu_{\Xi^-}) = 1.52$$

Experimentally, $S_{CG} = 0.49 \pm 0.05 \mu_N$.[4] This provides vital clues for studying SU(3) symmetry breaking effects.